\documentclass{PoS}

\def\beq#1{\begin{equation}\label{#1}}
\def\eeq{\end{equation}}
\def\beqa#1{\begin{eqnarray}\label{#1}}
\def\eeqa{\end{eqnarray}}

\def\Eq#1{Eq.~(\ref{#1})}

\def\myfrac#1#2{\left(\frac{#1}{#2}\right)}

\def\mycomment#1{\relax}

\title{Origin of bright flares in SFXTs}

\ShortTitle{Origin of bright flares in SFXTs}

\author{\speaker{Konstantin Postnov}, {Nikolai Shakura}\\
        Sternberg Astronomical Institute, Moscow M.V. Lomonosov University, Russia\\
        E-mail: \email{pk@sai.msu.ru}}

\author{Lara Sidoli, Ada Paizis\\
        INAF, Istituto di Astrofisica Spaziale e Fisica Cosmica, Milano,  Italy\\
        }

\abstract{In the settling accretion theory, which is applicable to 
quasi-spherical accreting slowly rotating 
magnetized neutron stars with X-ray luminosity $L_x\lesssim 
4\times 10^{36}$~erg/s, 
bright X-ray flares ($\sim 10^{38}-10^{40}$~ergs) observed in supergiant fast X-ray transients (SFXT)  may be produced by 
sporadic capture of magnetized stellar-wind plasma from the early-type supergiant. At sufficiently low steady accretion rates ($\lesssim 10^{15}$~g/s) through the shell around the neutron star magnetosphere at the settling accretion stage, 
magnetic reconnection 
can temporarily enhance the magnetospheric plasma entry rate, resulting in copious production of X-ray photons,
strong Compton cooling, and ultimately in unstable 
accretion of the entire shell. 
A bright flare develops on the free-fall time
scale in the shell, $R_B^{3/2}/\sqrt{GM}\sim 10^3-10^4$~s ($R_B$ is the 
classical Bondi capture radius), and the typical energy released in an SFXT bright flare corresponds to 
the mass of the shell.
}

\FullConference{10th INTEGRAL Workshop: A Synergistic View of the High-Energy Sky\\
                 15-19 September 2014\\
                 Annapolis, MD, USA}

\begin{document}

\section{Settling quasi-spherical accretion}

In close binary systems, 
there can be two different regimes of accretion onto the compact object -- disk accretion 
\cite{1973SvA....16..756S,1972A&A....21....1P,1973A&A....24..337S}
and quasi-spherical accretion. The disk accretion regime is usually takes place 
when the optical star overfills its Roche lobe. Quasi-spherical accretion is 
most likely to occur in high-mass X-ray binaries (HMXB) 
when an optical star of early spectral class (O-B) does not fill its Roche lobe, 
but experiences a significant mass loss via its stellar wind. We shall discuss the wind accretion regime,
in which a bow shock forms in the stellar wind around the compact star. The structure of the bow 
shock and the associated accretion wake is non-stationary and quite complicated (see e.g.
numerical simulations \cite{1988ApJ...335..862F, 1999A&A...346..861R, 2004A&A...419..335N}, 
among many others). The characteristic distance at which the bow shock forms
is approximately that of the Bondi radius $R_B=2GM/(v_w^2+v_{orb}^2)$, where $v_w$ is the wind velocity 
(typically 100-1000 km/s) and $v_{orb}$ is the orbital velocity of the compact star. In HMXBs, the stellar wind velocity is usually much larger than $v_{orb}$, so below we will neglect $v_{orb}$. 
The rate of gravitational capture of mass from a wind with density $\rho_w$
near the orbital position of the NS  
is the Bondi mass accretion rate: $\dot M_B\simeq \rho_w R_B^2 v_w\propto \rho_w v_w^{-3}$.   

Then, there are two different cases of quasi-spherical accretion. Classical 
Bondi-Hoyle-Littleton 
accretion takes place when the shocked matter is cooled down rapidly, 
and the matter falls freely towards the NS magnetosphere
by forming 
a shock at some distance above the magnetosphere. 
Here the shocked matter cools down (mainly via Compton processes) 
and enters the magnetopshere due to the Rayleigh-Taylor instability \cite{1976ApJ...207..914A}.
The magnetospheric boundary is characterized by the Alfv\'en radius $R_A$, which can be
calculated from the balance between the ram pressure of the 
infalling matter and the magnetic field pressure at the 
magnetospheric boundary. 
The captured matter from the wind carries a specific angular momentum 
$j_w\sim \omega_BR_B^2$ \cite{1975A&A....39..185I}. 
Depending on the sign of $j_w$ (prograde or retorgrade), the NS can spin-up 
or spin-down. This regime of quasi-spherical accretion occurs in 
bright X-ray pulsars with $L_x>4\times 10^{36}$~erg~s$^{-1}$ \cite{1983ApJ...266..175B, 2012MNRAS.420..216S}. 

If the captured wind matter behind the bow shock at $R_B$ remains hot (which it does when 
the plasma cooling time is much longer than the free-fall time, $t_{cool}\gg t_{ff}$), 
a hot quasi-static shell forms around the magnetosphere and subsonic 
(settling) accretion sets in. 
In this case, both spin-up and spin-down of the NS 
is possible, even if the sign of $j_w$ is positive (prograde). The shell mediates the
angular momentum transfer from the NS magnetosphere via viscous stresses
due to convection and turbulence. In this regime, the mean radial velocity 
of matter in the shell $u_r$ is smaller than the free-fall velocity $u_{ff}$: 
$u_r=f(u)u_{ff}$, $f(u)<1$, and is determined by the palsma cooling rate 
near the magnetosphere (due to Compton or radiative cooling): 
$f(u)\sim [t_{ff}(R_A)/t_{cool}(R_A)]^{1/3}$. In the settling accretion regime 
the actual mass accretion rate
onto the NS may be significantly smaller than the Bondi mass accretion rate, $\dot M=f(u) \dot M_B$.  
Settling accretion occurs 
at $L_x<4\times 10^{36}$~erg~s$^{-1}$ \cite{2012MNRAS.420..216S}. 

\subsection{Two regimes of plasma entering the NS magnetosphere}

To enter the magnetosphere, the plasma in the shell must cool down from a  
high (almost virial) temperature $T$ determined by hydrostatic equilibrium 
to some critical temperature $T_{cr}$ \cite{1977ApJ...215..897E} 
\beq{30}
{\cal R}T_{cr}=\frac{1}{2}\frac{\cos\chi}{\kappa R_A}\frac{\mu_mGM}{R_A}
\eeq
Here ${\cal R}$ is the universal gas constant, $\mu_m\approx 0.6$ is the molecular weight,
$G$ is the Newtonian gravitational constant, $M$ is the neutron star mass,  
$\kappa$ is the local curvature of the magnetosphere and $\chi$ is the angle 
between the outer normal and the radius-vector at any given point at the Alfv\'en surface.

As was shown in \cite{2012MNRAS.420..216S, 2014EPJWC..6402001S}, a transition zone above the 
Alfv\'en surface with radius $R_A$ is formed inside which the plasma cools down. The effective gravitational acceleration in this zone is 
\beq{}
g_{eff}=\frac{GM}{R_A^2}\cos\chi \left(1-\frac{T}{T_{cr}}\right)
\eeq 
and the mean radial velocity 
of plasma settling is 
\beq{}
u_R=f(u)\sqrt{2GM/R_A}\,.
\eeq
In the steady state, the dimensionless factor  
$0\le f(u)\le 1$ is determined by the specific plasma cooling mechanism 
in this zone and, by conservation of mass, is constant through the shell.  
This factor can be expressed through 
the plasma cooling time $t_{cool}$ in the transition zone \cite{2012MNRAS.420..216S}:
\beq{fu}
f(u) \simeq \myfrac{t_{ff}}{t_{cool}}^{1/3}\cos\chi^{1/3}
\eeq
where $t_{ff}=R^{3/2}/\sqrt{2GM}$ is the characteristic free-fall time scale from radius $R$. The angle
$\chi$ is determined by the shape of the magnetosphere, and for the magnetospheric boundary 
parametrized in the form   
$\sim \cos\lambda^n$ (where $\lambda$ is the angle counted from the magnetospheric equator) 
$\tan \chi=n\tan\lambda$. For example, in model calculations by \cite{1976ApJ...207..914A} $n\simeq 0.27$ in the near-equatorial zone, so $\kappa R_A\approx 1.27$.
We see that $\cos\chi\simeq 1$ up to $\lambda\sim \pi/2$, 
so below (as in \cite{2012MNRAS.420..216S}) we shall omit $\cos\chi$. 

Along with the density of 
matter near the magnetospheric boundary $\rho(R_A)$, 
the factor $f(u)$ determines the magnetosphere 
mass loading rate through 
the mass continuity equation:
\beq{cont}
\dot M=4\pi R_A^2 \rho(R_A) f(u)\sqrt{2GM/R_A}\,.
\eeq
This plasma eventually reaches the neutron star surface and 
produces an X-ray luminosity $L_x\approx 0.1\dot M c^2$.
Below we shall normalize the mass accretion rate 
through the magnetosphere as well as the X-ray luminosity 
to the fiducial values $\dot M_{n}\equiv \dot M/10^{n}$~g~s$^{-1}$ 
and $L_{n}\equiv L_x/10^{n}$~erg~s$^{-1}$, respectively.

\subsection{The Compton cooling regime}

As explained in detail in \cite{2012MNRAS.420..216S} (Appendix C and D), in subsonic quasi-static shells above slowly rotating NS magnetospheres 
the adiabaticity of the accreting matter is broken due to turbulent heating and Compton cooling. 
X-ray photons generated near the NS surface tend to cool down the matter in the shell via Compton scattering as long as the plasma
temperature $T>T_x$, where $T_x$ is the characteristic radiation temperature determined
by the spectral energy distribution of the X-ray radiation. 
For typical X-ray pulsars $T_x\sim 3-5$~keV. 
Cooling of the plasma at the base of the shell decreases the temperature gradient 
and hampers convective motions. Additional heating due to   
sheared convective motions is insignificant (see Appendix C of \cite{2012MNRAS.420..216S}).  
Therefore, the temperature in the shell 
changes with radius almost
adiabatically ${\cal R}T\sim (2/5)GM/R$, and 
the distance $R_x$ within which the plasma cools down 
by Compton scattering is
\beq{}
R_x\approx 10^{10}\hbox{cm}\myfrac{T_x}{3\hbox{keV}}^{-1}\,, 
\eeq
much larger than the characteristic Alfv\'en radius $R_A\simeq 10^9$~cm.

The Compton cooling time is inversely proportional to the 
photon energy density, 
\beq{t_C}
t_C\sim R^2/L_x\,, 
\eeq 
and near the Alfv\'en surface we find
\beq{t_Cn}
t_C\approx 10 \hbox{[s]} \myfrac{R_A}{10^9 \hbox{cm}}^2 
L_{36}^{-1}\,.
\eeq
(This estimate assumes spherical symmetry of the X-ray emission beam). 
Clearly, for the exact radiation density the shape of the X-ray emission  
produced in the accretion column near the NS surface (i.e. X-ray beam) 
is important, but still $L_x\sim \dot M$. 
Therefore, roughly, $f(u)_C\sim \dot M^{1/3}$, or, more precisely, 
taking into account the dependence of $R_A$ on $\dot M$, in this regime
\beq{RA_C}
R_A^C\approx 10^9\hbox{cm}L_{36}^{-2/11}\mu_{30}^{6/11}
\eeq
we obtain:
\beq{fu_C}
f(u)_C\approx 0.3L_{36}^{4/11}\mu_{30}^{-1/11}\,.
\eeq
Here $\mu_{30}=\mu/10^{30}$~G cm$^3$ is the NS dipole magnetic moment.

\subsection{The radiative cooling regime}

In the absence of a dense photon field, 
at the characteristic temperatures near the magnetosphere $T\sim 30$-keV and higher,
plasma cooling is essentially due to radiative losses (bremsstrahlung), and the plasma cooling time is 
$t_{rad}\sim \sqrt{T}/\rho$. Making use of the continuity equation (\ref{cont})
and the temperature distribution in the shell $T\sim 1/R$, we obtain 
\beq{t_rad}
t_{rad}\sim R \dot M^{-1} f(u)\,.
\eeq
Note that, unlike the Compton cooling time (\ref{t_C}), the radiative cooling time is 
actually independent of $\dot M$ (remember that $\dot M\sim f(u)$ in the subsonic accretion
regime!). 
Numerically, near the magnetosphere we have 
\beq{t_radn}
t_{rad}\approx 1000 \hbox{[s]} \myfrac{R_A}{10^9 \hbox{cm}}L_{36}^{-1}\myfrac{f(u)}{0.3}\,.
\eeq

Following the method described in Section 3 of \cite{2012MNRAS.420..216S}, we find the 
mean radial velocity of matter entering the NS magnetosphere in the 
near-equatorial region,
similar to the expression for $f(u)$ in the Compton cooling region \Eq{fu_C}. 
Using the expression for the Alfv\'en radius as expressed through $f(u)$, we calculate
the dimensionless settling velocity:  
\beq{fu_rad1}
f(u)_{rad}\approx 0.1 L_{36}^{2/9}\mu_{30}^{2/27}
\eeq 
and the Alfv\'en radius:
\beq{RA_rad}
R_A^{rad}\simeq 10^9\hbox{[cm]} L_{36}^{-2/9}\mu_{30}^{16/27}
\eeq
(in the numerical estimates we assume a monoatomic gas with adiabatic index 
$\gamma=5/3$). 
The obtained expression for the dimensionless settling velocity of matter 
\Eq{fu_rad1} in the radiative cooling regime clearly shows that here accretion proceeds
much less effectively than in the Compton cooling regime (cf. with \Eq{fu_C}).   

Unlike in the Compton cooling regime, 
in the radiative cooling regime there is no instability leading to an increase of the mass
accretion rate as the luminosity increases (due to the long characteristic
cooling time), 
and accretion here is therefore expected to proceed more quietly
under the same external conditions.

The idea that the transition between the two regimes may be triggered by a change in the X-ray beam pattern is supported by the pulse profile observations of Vela X-1 
in different energy bands \cite{Doroshenko_ea11}. 
The observed change in phase of the 20--60 keV profile in the off-state (at X-ray luminosity $\sim 2.4\times 10^{35}$~erg~s$^{-1}$), reported by \cite{Doroshenko_ea11}, suggests a disappearance of the fan beam at hard X-ray energies upon the source entering this state and the formation of
a pencil beam
(see \cite{2013MNRAS.428..670S} for more detailed discussion).

Note that the pulse profile phase change associated with X-ray beam switching below some critical luminosity, as observed in Vela X-1, seems to be suggested by an $XMM-Newton$ observation of the SFXT IGR~J11215--5952 (see Fig.~3 in \cite{Sidoli2007}),
corroborating the subsonic accretion regime with radiative plasma cooling at low X-ray luminosities in SFXTs as well, as we shall describe in the next section (see 
\cite{2014MNRAS.442.2325S} for more detailed discussion).

\section{SFXTs}

Supergiant Fast X-ray Transients (SFXTs) are a subclass of 
HMXBs 
associated with early-type supergiant companions \cite{Pellizza2006, Chaty2008, Rahoui2008},
and characterized by sporadic, short and bright X--ray flares 
reaching peak luminosities of 10$^{36}$--10$^{37}$~erg~s$^{-1}$.
Most of them were discovered by INTEGRAL \cite{2003ATel..176....1M, 2003ATel..190....1S, 
2003ATel..192....1G, Sguera2005, Negueruela2005}.
They show high dynamic ranges (between 100 and 10,000, depending on the specific source; 
e.g. \cite{Romano2011, 2014A&A...562A...2R}) and their X-ray spectra in outburst are very similar to accreting pulsars in HMXBs.  
In fact, half of them have measured neutron star (NS) spin periods similar to those observed 
from persistent HMXBs (see \cite{2012int..workE..11S} for a recent review).

The physical mechanism driving their transient behavior, related to the accretion by the compact
object of matter from the supergiant wind, has been discussed by several authors
and is still a matter of debate, as some of them require particular properties of the compact objects hosted in these systems 
\cite{2007AstL...33..149G, 2008ApJ...683.1031B}, 
and others assume
peculiar clumpy properties of the supergiant winds and/or orbital characteristics 
\cite{zand2005,Walter2007,
Sidoli2007,
Negueruela2008,2009MNRAS.398.2152D,Oskinova2012}.

\textbf{Energy released in bright flares.}
The typical energy released in a SFXT bright flare is about 
$10^{38}-10^{40}$~ergs \cite{2014arXiv1405.5707S}, 
varying by one order of magnitude between different
sources. That is, the mass fallen onto the NS
in a typical bright flare varies from $10^{18}$~g to around $10^{20}$~g. 

The typical X-ray luminosity outside outbursts in SFXTs is about 
$L_{x,low}\simeq 10^{34}$~erg s$^{-1}$ \cite{2008ApJ...687.1230S},
 and below we shall normalise the luminosity to this value, $L_{34}$. 
At these low X-ray luminosities, the plasma entry rate into the magnetosphere is controlled 
by radiative plasma cooling.
Further, it is convenient to normalise the typical stellar wind velocity from hot OB-supergiants $v_w$ 
to 1000~km~s$^{-1}$ (for orbital periods of about a few days or larger the NS orbital velocities can be neglected compared to the stellar wind velocity from the OB-star), so that the Bondi gravitational capture radius is $R_B=2GM/v_w^2=4 \times 10^{10}v_{8}^{-2}$~cm 
for a fiducial NS mass of $M_x=1.5 M_\odot$.

Let us assume that a quasi-static shell hangs over the magnetosphere around the NS, with the magnetospheric accretion rate being controlled by radiative plasma cooling.
We denote the actual steady-state accretion rate as $\dot M_a$ so that the observed X-ray steady-state luminosity  is $L_x=0.1\dot M_a c^2$. Then from the theory of subsonic 
quasi-spherical accretion \cite{2012MNRAS.420..216S} we know that the factor $f(u)$ (the ratio of the actual velocity of plasma entering the magnetosphere, due to the Rayleigh-Taylor instability, 
to the free-fall velocity at the magnetosphere,
$u_{ff}(R_{A})=\sqrt{2GM/R_A}$) reads \cite{2013MNRAS.428..670S,2014EPJWC..6402001S}
\beq{furad}
f(u)_{rad} \simeq 0.036 L_{34}^{2/9}\mu_{30}^{2/27}\,.
\eeq
(See also Eq. (\ref{fu_rad1}) above).

The shell is quasi-static (and likely convective), unless something triggers a 
much faster matter fall through the magnetosphere (a possible reason is suggested below). 
It is straightforward to calculate the mass of the shell 
using the density distribution $\rho(R)\propto R^{-3/2}$ 
\cite{2012MNRAS.420..216S}. Using the mass continuity equation to eliminate the density above the magnetosphere, we readily find 
\beq{deltaM}
\Delta M \approx \frac{2}{3} \frac{\dot M_a}{f(u)}t_{ff}(R_B)\,.
\eeq
Note that this mass can be expressed through measurable quantities
$L_{x,low}$, $\mu_{30}$ and the (not directly observed) stellar wind
velocity at the Bondi radius $v_w(R_B)$. Using Eq. (\ref{furad}) for  
the radiative plasma cooling, we obtain
\beq{deltaMrad}
\Delta M_{rad} 
\approx 8\times 10^{17} [g] L_{34}^{7/9} v_8^{-3}\mu_{30}^{-2/27}\,.
\eeq
The simple estimate (\ref{deltaMrad}) shows that for a typical wind velocity 
near the NS of about 500~km~s$^{-1}$ the \emph{typical} mass of the hot 
magnetospheric shell is around $10^{19}$~g, 
corresponding to $10^{39}$~ergs released in a flare in which all the matter from the shell
is accreted onto the NS, as observed. 
Clearly, variations in stellar wind velocity between different sources by a factor of $\sim 2$  
would produce the one-order-of-magnitude spread in $\Delta M$ observed in bright SFXT flares.

\begin{figure*}
\includegraphics[width=0.7\textwidth]{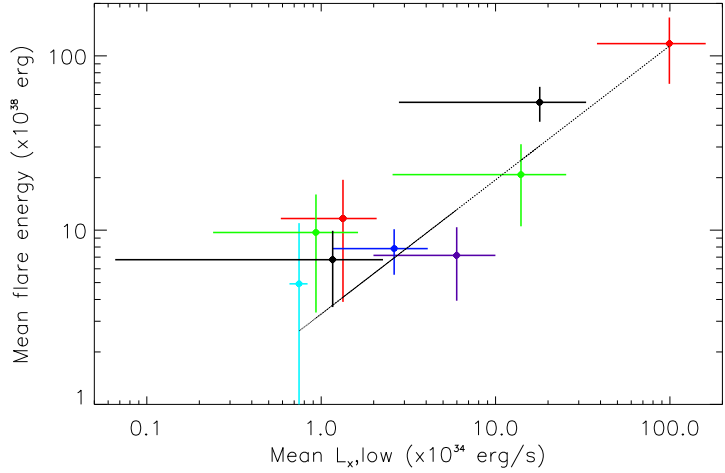}
\caption{The mean energy released in bright flares (17 -- 50 keV, data
from \cite{PaizisSidoli2014}) versus average
\textit{INTEGRAL}/IBIS source luminosity.
The \textit{x-axis}
is in units of $10^{34}$~erg~s$^{-1}$,
the \textit{y-axis} is in units of $10^{38}$~ergs.
The straight line gives the formal rms linear fit with the slope 
$0.77 \pm 0.13$. (Figure adapted from \cite{2014arXiv1405.5707S}). }
\label{fig:dM}
\end{figure*}

In Fig. \ref{fig:dM} we show the mean energy of SFXT bright flares 
$\Delta E=0.1\Delta M c^2$ as a function of the low (non flaring) X-ray luminosity 
for nine SFTXs from our recent paper \cite{2014arXiv1405.5707S}. 
The low (non flaring) X-ray luminosity (\emph{x-axis}) has been taken from 
\cite{krivonos2012}, where a nine year time-averaged source flux in the 17--60\,keV band is given for each source \footnote{IGR\,J17544--2619, IGR\,J16418--4532, IGR\,J16479--4514, IGR\,J16465--4507, SAX\,J1818.6--1703,  IGR\,J18483--0311, XTE\,J1739--302, IGR\,J08408--4503, IGR\,J18450--0435, IGR\,J18410--0535, IGR\,J11215--5952}. The data selection and analysis is discussed in detail in \cite{PaizisSidoli2014}, together with the assumed distances and relevant references, so we refer the reader to that paper for the technical details.
The uncertainties on the low luminosities  
include both the statistical errors on source fluxes, as reported in \cite{krivonos2012},
and the known SFXTs distances and their uncertainties as reported by  \cite{PaizisSidoli2014}. 
The formal rms fit to these points, shown by the straight line, gives the dependence of 
$\Delta E_{38}=(3.3\pm 1.0)L_{34}^{0.77\pm  0.13}$. 
This exactly corresponds to the radiative cooling regime $\Delta E\propto L^{7/9}$ 
(see Eq. (\ref{deltaMrad})), as expected.  
A comparison with the coefficient in expression (\ref{deltaMrad}) suggests $v_8\sim 0.62$,  
similar to typical wind velocities observed in HMXBs.

What can trigger SFXT flaring activity? As noted in 
\cite{2013MNRAS.428..670S},
if there is an instability leading to a rapid fall of matter through the magnetosphere, 
a large quantity of X-ray photons produced near the NS surface should 
rapidly cool down the plasma near the magnetosphere, further increasing the plasma fall velocity
$u_R(R_A)$ 
and the ensuing accretion NS luminosity $L_x$. Therefore, in a bright flare 
the entire shell can fall onto the NS on the free-fall time scale from the outer 
radius of the shell $t_{ff}(R_B)\sim 1000$~s. Clearly, the shell will be replenished by
new wind capture, so the flares will repeat as long as the 
rapid mass entry rate into the magnetosphere is sustained.

\vskip\baselineskip
\noindent
\textbf{Magnetized stellar wind as the flare trigger.}
We suggest that the shell instability described above can be 
triggered by a large-scale magnetic field sporadically 
carried by the stellar wind of the optical OB 
companion. Observations suggest that about $\sim 10\%$ of hot OB-stars have magnetic fields
up to a few kG (see \cite{2013arXiv1312.4755B} for a recent review and discussion).
It is also well known from Solar wind studies (see e.g. reviews \cite{2004PhyU...47R...1Z, lrsp-2013-2} and references therein) that the Solar wind patches carrying tangent magnetic fields 
has a  lower velocity (about $350$~km~s$^{-1}$) than the wind with radial magnetic fields 
(up to $\sim 700$~km s$^{-1}$). Fluctuations of the stellar wind density and velocity 
from massive stars are also known from spectroscopic observations \cite{2008A&ARv..16..209P}, 
with typical velocity fluctuations up to $0.1\ v_\infty\sim 200-300$~km s$^{-1}$.  

The effect of the magnetic field carried by the stellar wind is twofold: first, 
it may trigger
rapid mass entry to the magnetosphere via magnetic reconnection in the magnetopause (the phenomenon well known in the dayside Earth magnetosphere, \cite{1961PhRvL...6...47D}), and secondly, 
the magnetized parts of the wind (magnetized clumps with a tangent magnetic field) have a lower velocity than the non magnetised 
ones (or the ones carrying the radial field). As discussed in \cite{2014arXiv1405.5707S} and below,
magnetic reconnection 
can increase the plasma fall velocity in the shell from inefficient, radiative-cooling controlled settling accretion 
with $f(u)_{rad}\sim 0.03-0.1$, 
up to the maximum possible free-fall velocity with $f(u)=1$.
In other words, during a bright flare subsonic 
settling accretion turns into supersonic Bondi accretion.
The second factor (slower wind velocity in magnetized clumps with tangent magnetic field) 
strongly increases the Bondi radius $R_B\propto v_w^{-2}$
and the corresponding Bondi mass accretion rate $\dot M_B\propto v_w^{-3}$. 

Indeed, we can write down the mass accretion rate onto the NS in the unflaring
(low-luminosity) state as $\dot M_{a,low}=f(u) \dot M_B$
with $f(u)$ given by expression (\ref{furad}) 
and $\dot M_B\simeq \pi R_B^2 \rho_w v_w $.
Eliminating the wind density $\rho_w$ using the mass continuity equation written for the 
spherically symmetric stellar wind from the optical star with power $\dot M_o$ and assuming  
a circular binary orbit, we arrive at 
$
\dot M_B\simeq \frac{1}{4}\dot M_o \myfrac{R_B}{a}^2\,.
$
Next, let us utilize 
the well-known relation for the radiative wind mass-loss rate from massive hot stars
$
\dot M_o\simeq \epsilon \frac{L}{cv_\infty}
$
where $L$ is the optical star luminosity, $v_\infty$ is the stellar wind velocity at infinity,
typically 2000-3000 km s$^{-1}$ for OB stars and $\epsilon\simeq 0.4-1$ is the efficiency factor \cite{1976A&A....49..327L} (in the numerical estimates below we shall assume $\epsilon=0.5$). 
It is also possible to reduce the luminosity $L$ of a
massive star to its mass $M$ using 
the phenomenological relation $(L/L_\odot)\approx 19 (M/M_\odot)^{2.76}$ (see e.g. \cite{2007AstL...33..251V}). Combining the above equations and using
Kepler's third law to express the orbital separation $a$ through the binary period $P_b$, we find for the X-ray luminosity of SFXTs in the non-flaring state 
\begin{eqnarray}
\label{Lxlow}
L_{x,low}\simeq & 5\times 10^{35} [\hbox{erg~s}^{-1}] f(u) 
\myfrac{M}{10 M_\odot}^{2.76-2/3} \nonumber\\
&\myfrac{v_\infty}{1000 \mathrm{km~s}^{-1}}^{-1}
\myfrac{v_w}{500 \mathrm{km~s}^{-1}}^{-4}\myfrac{P_b}{10 \mathrm{d}}^{-4/3}\,,
\end{eqnarray}
which for $f(u)\sim 0.03-0.1$ 
corresponds to the typical low-state luminosities of SFXTs of $\sim 10^{34}$~erg~s$^{-1}$. 

It is straightforward to see that a transition from the low state (subsonic accretion with 
slow magnetospheric entry rate $f(u)\sim 0.03-0.1$) to supersonic free-fall Bondi accretion 
with $f(u)=1$ due to the magnetized stellar wind with the velocity decreasing by a factor of two, for example, would lead to a flaring luminosity of $L_{x,flare}\sim (10\div 30)\times 2^5 L_{x,low}$. This shows that 
the dynamical range of SFXT bright flares ($\sim 300-1000$) can be naturally reproduced by the proposed mechanism.

\textbf{Conditions for magnetic reconnection.}
For magnetic field reconnection to occur, the time the magnetized plasma spends 
near the magnetopause should be at least comparable to 
the reconnection time, $t_r\sim R_A/v_r$, where
$v_r$ is the magnetic reconnection rate, which is difficult to assess from first principles
\cite{2009ARA&A..47..291Z}.
For example, in the Petschek fast reconnection model $v_r=v_A(\pi/8\ln S)$, where $v_A$ is the 
Alfv\'en speed and $S$ is the Lundquist number (the ratio of the global Ohmic dissipation 
time to the Alfv\'en time); for typical conditions near NS magnetospheres we
find $S\sim 10^{28}$ and $v_r\sim 0.006 v_A$. In real astrophysical plasmas 
the large-scale magnetic reconnection rate can be a few times as high, 
$v_r\sim 0.03-0.07 v_A$ \cite{2009ARA&A..47..291Z}, and, guided by phenomenology, we can parametrize it as $v_r=\epsilon_r v_A$ with $\epsilon_r\sim 0.01-0.1$. The longest time-scale the plasma penetrating into the magnetosphere spends near the magnetopause
is the instability time, $t_{inst}\sim t_{ff}(R_A)f(u)_{rad}$ \cite{2012MNRAS.420..216S}, so the 
reconnection may occur if 
$t_r/t_{inst}\sim (u_{ff}/v_A)(f(u)_{rad}/\epsilon_r)\lesssim 1$. As near 
$R_A$ (from its definition) $v_A\sim u_{ff}$, we arrive at $f(u)_{rad}\lesssim\epsilon_r$ as
the necessary reconnection condition. According to Eq. (\ref{furad}), it is satisfied only 
at sufficiently low X-ray luminosities, pertinent to 'quiet' SFXT states. 
\textit{This explains why in HMXBs with convective shells at higher luminosity
(but still lower than $4\times 10^{36}$~erg~s$^{-1}$, at which settling accretion is possible), 
%
reconnection from magnetised plasma accretion will not
lead to shell instability, but only 
to temporal establishment of the 'strong coupling regime' 
of angular momentum transfer through the shell, as 
discussed in \cite{2012MNRAS.420..216S}.} 
Episodic strong spin-ups, as observed in GX 301-2, 
may be manifestations of such 'failed' 
reconnection-induced shell instability.

Therefore, it seems likely that the key difference between 
steady HMXBs like Vela X-1, GX 301-2 (showing only moderate flaring activity) and SFXTs is
that in the first case the effects of possibly magnetized stellar winds from optical OB-companions
are insignificant (basically due to the rather high mean accretion rate),
while in SFXTs with lower 'steady' X-ray luminosity, 
large-scale magnetic fields, sporadically carried by clumps in the wind, 
can trigger SFXT flaring activity via magnetic reconnection near the magnetospheric boundary. 
The observed power-law SFXT flare distributions, discussed in \cite{PaizisSidoli2014},
with respect to the log-normal distributions for classical HMXBs \cite{2010A&A...519A..37F}, 
may be related to the properties of magnetized stellar wind and physics of its interaction 
with the NS magnetosphere.

\textbf{Acknowledgement.} {The work of KP and NSh is supported by the Russian Fund for Basic Research through grant 14-02-00657a. LS and AP acknowledge the Italian Space Agency financial support INTEGRAL ASI/INAF agreement n. 2013-025.R.0. }

\bibliographystyle{azh}
\bibliography{wind_z}


\end{document}